\documentstyle[12pt]{article}
\newcommand{\rhohat}{{\hat{\rho}(1600)}}

\newcommand{\btab}{\begin{tabbing}}
\newcommand{\etab}{\end{tabbing}}

\newcommand{\beqn}{\begin{equation}}
\newcommand{\eeqn}{\end{equation}}
\newcommand{\barr}[1]{\begin{array}{#1}}
\newcommand{\earr}{\end{array}}
\newcommand{\beqna}{\begin{eqnarray}}
\newcommand{\eeqna}{\end{eqnarray}}
\newcommand{\btablec}{\begin{table} \begin{center}}
\newcommand{\etablec}{\end{center} \end{table}}
\newcommand{\lapprox}{\stackrel{<}{\scriptstyle \sim}}

\newcommand{\gapproxeq}{\lower.7ex\hbox{$\;\stackrel{\textstyle>}{\sim}\;$}}
\newcommand{\plabel}[1]{\label{#1}}
\newcommand{\pbibitem}[1]{\bibitem{#1}}
\input epsf

\begin{document}
\title{\begin{flushright} \small{hep-ph/9709231}\\
\small{JLAB-THY-97-37} \end{flushright} 
\vspace{0.6cm}  
\LARGE \bf Implications of a $J^{PC}$ exotic}
\author{Philip R. Page\thanks{\small \em E-mail:
prp@jlab.org. Tel: +1-757-2695028.  Fax: +1-757-2697002.} \\
{\small \em Theory Group, Thomas Jefferson National Accelerator Facility,}\\ 
{\small \em 12000 Jefferson Avenue, Newport News, VA 23606, USA.}}
\date{September 1997}
\maketitle
\abstract{Recent experimental data from BNL on the isovector
$J^{PC}=1^{-+}$ exotic at 1.6 GeV in $\rho\pi$
indicate the existence of a non--quarkonium state consistent with
lattice gauge theory predictions. We discuss how further experiments
can strengthen this conclusion. We show that the $\rho\pi,\;
\eta^{'}\pi$ and $\eta\pi$ couplings of this state qualitatively support the
hypothesis that it is a hybrid meson, although other interpretations
cannot be eliminated.}

\vspace{2cm}

\normalsize

There are two independent indications of an
isovector $J^{PC}=1^{-+}$ exotic resonance $\rhohat$ in $\pi^{-} N \rightarrow
\pi^{+}\pi^{-}\pi^{-} N$.
The E852 collaboration at BNL recently reported evidence for a resonance at
$1593\pm 8$ MeV with a width of
$168\pm 20$ MeV \cite{bnl97}.
These parameters are consistent with
the preliminary claim by the VES collaboration of a resonance at $1.62\pm 0.02$ GeV with a width of
$0.24\pm 0.05$ GeV \cite{ves93}. In both cases a partial wave analysis
was performed, and the decay mode $\rho^{0}\pi^{-}$ was observed.

In this letter we indicate that experimental and lattice gauge theory
$1^{-+}$ mass estimates are converging. To
further verify the experimental signal and to provide independent confirmation, we suggest a search in other
decay channels and for other $J^{PC}$ exotic resonances. If 
current experimental data are taken at face value, we show that the $\rho\pi,\;
\eta^{'}\pi$ and $\eta\pi$ couplings of $\rhohat$ qualitatively support the
hypothesis that it is a hybrid meson, although other interpretations
cannot be eliminated. We indicate further production mechanisms.

There are several reasons to believe that the resonance at 1.6 GeV in
$\rho\pi$ is not an experimental artifact: 

\begin{enumerate}


\item Although the $1^{-+}$ signal is weak, the well--known states
$\pi(1300)$ and $\pi(1800)$ are of similar intensity and are clearly
seen, suggesting that the exotic signal should be taken seriously \cite{bnl97,ves93}.

\item An extensive study was done by E852 to eliminate any sources of 
spurious signals. Instrumental acceptance and smearing of angular distributions (called
``leakage'') have been analysed, but the $J^{PC}=1^{-+}$ peak at 1.6
GeV appears to be robust. The fact that $\rhohat$ survives the leakage studies is
non--trivial, given that the exotic signal below 1.5 GeV was shown to
be an artifact of leakage. There is {\it a priori} the possibility that the only known state with a mass in the range 1.5 -- 1.7
GeV which can contribute (within isospin symmetry) in the
$\pi^{+}\pi^{-}\pi^{-}$ channel, the $\pi_2(1670)$,  can appear as a $1^{-+}$ signal due to leakage.
E852 finds the $\pi_2(1670)$ mass to be is $1683\pm 4$ MeV \cite{manak},
which is not only very precise but also consistent with the PDG value
$1670\pm 20$ MeV \cite{pdg96}. The width measured by E852 is $277\pm
10$ MeV, consistent with the PDG value $258\pm 18$ MeV. These 
masses and widths are very different from those measured for the exotic signal.

\item E852 observes rapid phase motion in the $1.5-1.7$ GeV region relative to the
$0^{-+},\;2^{++},$ $1^{++}$ and $2^{-+}$ partial waves.
It is impressive that clear phase motion is observed relative to
the nearby $\pi_2(1670)$. VES preliminarily indicated proper phase variation 
with respect to the $2^{++}$ wave, although the details of the phase
motion is somewhat different \cite{ves93}. The overall conclusion is that
$\rhohat$ is resonant.

\item The preliminary E852 analysis of $\pi^{-} N \rightarrow
\pi^{-}\pi^{0}\pi^{0} N$ is consistent with most of the features
in the $\pi^{-} N \rightarrow \pi^{+}\pi^{-}\pi^{-} N$ analysis \cite{brown}. 


\end{enumerate}

E852 showed that the $\rhohat$ is produced slightly more strongly
in cases where the incoming pion beam couples to the exchanged $b_1$ or
$f_1$ Regge trajectories, rather than the $\rho$ or $f_2$ trajectories \cite{bnl97}.
This suggests that exchange of states more massive than 1 GeV
can be important for production of $\rhohat$.

Further indications that there may be an exotic at 1.6 GeV are:

\begin{enumerate}

\item VES reported a broad structure peaking at $\sim 1.6$ GeV
(recently estimated to be $1595\pm 17\pm 10$ MeV \cite{zaitsev97}) with considerably stronger coupling to $\eta^{'}\pi$ than $\eta\pi$ for invariant
masses greater than $1.4 $ GeV when the effects 
of phase space have been removed \cite{ves93}. There is no phase
motion against the $a_2$, indicating that it is non--resonant. 

\item BNL and VES reported a $1^{-+}$ exotic wave in $f_1\pi$, 
which is unlikely to arise from misidentification 
of the dominant $1^{++}$ structure due to the similar intensity of the
$1^{-+}$ and $1^{++}$ waves \cite{lee94,todd,ves}. 
However, it is experimentally challenging
to reconstruct the $f_1$ and the results should hence be regarded with
some caution as ``more data would be required for a firm conclusion''
\cite{lee94}. Particularly, we note that $\sim 500$ events per 100
MeV bin were collected by BNL and VES in the $f_1\pi$ channel
\cite{lee94,ves}, compared with
$\sim 10000$ events per 100
MeV bin in $\rho\pi$ \cite{bnl97}. The $K^+\bar{K}^0\pi^-\pi^-$ channel analysed
by BNL yielded a structure at $1.7$ GeV which decayed
to $f_1\pi$.
However, they report no phase motion relative to the dominant $1^{++}$
structure, indicating that it is non--resonant if the $1^{++}$ is
non--resonant. This is confirmed by 
recent data from BNL which shows $1^{-+}$ peaking at $1.7-1.8$ GeV
in $f_1\pi\rightarrow\eta\pi^+\pi^-\pi^-$ \cite{todd}, while VES finds a peak
at $1.7-1.8$ GeV in the same channel \cite{ves}.  

\end{enumerate}

A second, independent, reason to support an exotic signal at greater
than 1.5 GeV arises from narrowing estimates for $1^{-+}$ masses from lattice gauge theory.
There are two groups that have predicted $s\bar{s}$ $1^{-+}$,
the UKQCD  \cite{lacock96} and MILC collaborations \cite{degrand96,degrand97}.
These groups have considerable facility to separate the lightest
$1^{-+}$ state for a given quark mass. MILC has presented evidence that
operators, whether they are hybrid--like or four--quark like,
easily select the lowest--lying state, indicating a strong coupling of
either operator to the $1^{-+}$ state. It is important to
emphasize that quenched lattice calculations allow Fock state
components with $Q\bar{Q}$ and $Q\bar{Q}Q\bar{Q}$ quark structure, and that the quark Fock state decomposition
of the lowest--lying $1^{-+}$ is by no means determined by these
calculations.  Although lattice calculations in the quenched
approximation contain systematic errors
due to contamination from
higher excited states 
and lattice artifacts, they
are settling down on an $s\bar{s}$ mass estimate. UKQCD
estimates $1990\pm 130$ MeV, where the error is
statistical. MILC quotes $2170\pm 80$ MeV 
with an additional $\sim 200$ MeV systematic error. 
Systematic errors do not include errors due to the quenched
approximation. Taking the mass difference between light and $s\bar{s}$
hybrids to be similar to conventional mesons, we expect light quark
hybrids to be $\sim 200-250$ MeV lighter \cite{pdg96}, i.e. $(1.75-2)\pm 0.2$ GeV. In fact,
MILC finds $1970\pm 90$ MeV from an extrapolation of hybrid
masses to light quarks, with a systematic error of 300 MeV.

The existence of $\rhohat$ may be confirmed by searching for its
presence in hithero unexplored decay channels like $K^{\ast}K$, but
particularly\footnote{$b_1\pi$ is the dominant mode predicted in
flux--tube models \protect\cite{models}.} in $b_1\pi$. A $b_1\pi$ analysis is in progress
at BNL and has provisionally been approved at TJNAF \cite{cebafb1pi}.

Another way to confirm the isovector $1^{-+}$ at 1.6 GeV is by searching for 
its non--$s\bar{s}$ isoscalar partner in the same mass region. 
Decays to the final states $\eta^{'}\eta,\; \rho\rho$ and
$\omega\omega$ are expected\footnote{$0^{+-},\; 1^{-+}$ and
$2^{+-}\rightarrow\pi^0\pi^0, \pi^+\pi^-,\; K^+K^-,\; K^0\bar{K}^0,\;\eta\eta$ and
$\eta^{'}\eta^{'}$ are forbidden by $CP$ conservation and Bose symmetry.}, although the
first mode is suppressed by symmetrization
selection rules \cite{page97sel1,tech}. Null results in
$ \eta^{'}\eta$ should confirm the selection rule.
$K^{\ast}K$ and $a_2\pi$ 
are also search channels, and a study of
$K^{\ast}K$ is already under way at BNL, although nothing has been reported
in this channel by VES \cite{ves}. Additional channels\footnote{Dominant channels for
flux--tube model hybrids \protect\cite{models}.}
are $\pi(1300)\pi$ and $a_1\pi$, but these involve
broad final states. From the decay modes listed it follows that production via incoming pions\footnote{Throughout this
work we shall assume that s--channel and u--channel production of states in the mass
range of interest are suppressed, since very heavy $\approx 3$ GeV 
excited nucleons would have to be produced for this mechanism
to be viable.} at BNL and VES can only proceed via
exchange of states more massive than 1 GeV. Photoproduction at TJNAF can
occur via $\rho$ exchange.
 
The existence of $\rhohat$ can be confirmed by searching for the
$J^{PC}$ exotic partners $0^{+-}, 2^{+-}$ and $0^{--}$. 
 Lattice gauge theory estimates for their $s\bar{s}$ masses exist.
UKQCD estimates that the $0^{+-}$ and $2^{+-}$ are 
respectively $0.2\pm 0.2$ GeV and $0.7\pm 0.3$ GeV more massive
than the $1^{-+}$ \cite{lacock96}. MILC remarks that $0^{+-}$ is ``clearly heavier than the $1^{-+}$''
and estimates the mass difference to be $0.2\pm 0.1$ GeV
(for $c\bar{c}$) \cite{degrand97}. The message appears to be that the $0^{+-}$ is only
slightly heavier than the $1^{-+}$.    
No firm numerical estimate exists for the $0^{--}$ exotic,
suggesting that if it exists at all, it is very heavy \cite{degrand97}. 

The isovector $0^{+-}$ can decay to $\pi\pi$ and $\rho\rho$,
but these decays are suppressed by symmetrization selection rules \cite{tech}.
$b_1\eta$ and $K_1(1270)K$ are possible, but may be suppressed by P--wave phase space. 
Decays to $a_1\pi,\; h_1\pi$ and $\pi(1300)\pi$ involve
broad final states\footnote{Dominant channels for
flux--tube model hybrids \protect\cite{models}.}.
The isovector $0^{+-}$
may hence be very challenging to isolate experimentally due to its
lack of coupling to tractable final states \cite{models}.

The isoscalar $0^{+-}$ can decay to $K_1(1270)K$ and $h_1\eta$, but both are likely
to be suppressed by P--wave phase space. The {\it only} other strong
decay mode that is expected to be significant is $b_1\pi$. This leads
to the extraordinary prediction that the isoscalar $0^{+-}$
essentially only couples to $b_1\pi$. This mode should be
allowed in future $b_1\pi$ analyses.
  
For both the isovector and isoscalar $0^{+-}$ an incoming pion beam at VES and BNL 
can couple only to exchanged mesons more massive than 1
GeV. In the case of the isoscalar this is also true for
a photon beam at TJNAF. However, the isovector can be photoproduced
via $\rho$ exchange.

Lastly we consider the $2^{+-}$ states. These are expected to be more
massive than the $1^{-+}$ and $0^{+-}$ exotics, and have more decay channels than
the $0^{+-}$ states even at the same mass. We hence {\it a priori}
expect these states to be wider\footnote{The total width
of $2^{+-}$ hybrids are anomalously small in the PSS flux--tube model
\protect\cite{models}.}. This disadvantage is partially offset by the fact
that these states can decay to final states which are easy to access
experimentally, some of which we list here. Particularly, the
isovector can decay to $\omega\pi,\; K^{\ast}K,\; \rho\eta$ and $\rho\rho$. Note that decay to
$\pi\pi$ is suppressed by symmetrization selection
rules \cite{tech}. The isoscalar couples to $\rho\pi,\; K^{\ast}K$ and
$\omega\eta$. The allowed decays listed are suppressed D--wave phase space. 
The $2^{+-}$ states can be produced at VES and BNL via
vector meson exhange and at TJNAF via $\pi$ and $\eta$ exchange (and
$\rho$ exchange for the isovector). Additional
modes to narrow final states are $K_1(1270)K$, as well as isovector
$2^{+-}$ decay to $a_2\pi$ and isoscalar
decay to $b_1\pi$. 

We proceed to estimate the $\rho\pi$ coupling of $\rhohat$.
Estimates for $\rho\pi$ couplings can be obtained by noticing that
the E852 pion beam experiment established that $a_2,\; a_1$ and $\pi_2(1670)$
are produced via natural parity exchange, i.e. mainly the
$\rho$ trajectory\footnote{Exchange of the $f_2$ trajectory should be
suppressed on mass grounds, and since no significant $a_2,\;
a_1,\; \rhohat\rightarrow f_2\pi$ was reported
(although $\pi_2(1670)\rightarrow f_2\pi$ was observed) \protect\cite{bnl97}.
We also neglect diffractive exchange due to the low acceptance of the
experiment at small $t < 0.1$ GeV$^2$ \protect\cite{manak}, although
this possibility has not been eliminated experimentally.}. 
The fact that the states also decay to $\rho\pi$
enables us to make a rough phenomenological estimate of the partial
width of the states to $\rho\pi$. The number of events observed should
be proportional to the coupling of the incoming pion to the exchanged
$\rho$ and the probability of decay of the state $X$ to
$\rho\pi$, i.e. to $|{\cal M}|^2(X\rightarrow \rho\pi)\;
BR(X\rightarrow\rho\pi)$. The amplitude ${\cal M}$ and the width
$\Gamma$ are related by
$|{\cal M}|^2 = \frac{8\pi m_X^2}{|{\bf q}|} \; (2J_X+1)\; \Gamma$ \cite{pdg96}, where
$m_X$ is the mass of $X$, ${\bf q}$ the momentum of $\rho$ in the
rest frame of $X$ and $J_X$ the angular momentum of $X$.
Hence,

\beqn \plabel{wid}
\Gamma(X\rightarrow \rho\pi)\propto \sqrt{\mbox{(Number of events)}_X
\mbox{(Total width)}_X\; \frac{|{\bf q}|}{(2 J_X +1)\; m_X^2}}
\eeqn
We estimate\footnote{We estimate
100000, 370000, 60000 and 8000 events in natural parity exchange
under the $a_2,\; a_1,\; \pi(1670)$ and $\rhohat$ Breit--Wigner peaks. The total
widths used are respectively $107,\; 350,\; 258$ and $168$ MeV and the
$\rho\pi$ widths used are respectively $74,\; 350,\; 80$ MeV \protect\cite{bnl97,pdg96}.} $R(X) \equiv \Gamma(X\rightarrow \rho\pi)/\Gamma(a_2\rightarrow
\rho\pi)$ according to Eq. \ref{wid} from E852 data \cite{bnl97}

\beqn
R(a_1) \approx 4.4 \hspace{1cm}
R(\pi_2(1670)) \approx 1.2 \hspace{1cm}
R(\rhohat) = 0.4 - 0.5
\eeqn
while estimates from the PDG gives

\beqn
\hspace{-5.2cm} R(a_1) = 4.7 \hspace{1cm}
R(\pi_2(1670)) = 1.1
\eeqn
It is clear that the na\"{\i}ve estimates of $R(X)$ from E852
data are in accord
with expectations from the PDG, motivating the use of Eq. \ref{wid}. Since $R(\rhohat) = 0.4-0.5$ we can
predict that $\Gamma(\rhohat\rightarrow \rho\pi) = 30 - 37$ MeV
and $BR(\rhohat\rightarrow \rho\pi) = 20\pm 2\;\%$.  

An experimental limit on $BR(\rhohat^+\rightarrow
\rho^0\pi^+)\lapprox 20\%$ has been published for a $1^{-+}$ state 
at 1.5 GeV \cite{ziel}. This translates to $BR(\rhohat\rightarrow
\rho\pi)\lapprox 40\%$, well within the estimates presented above.
Since these limits are not available 
at 1.6 GeV, we assume that the results remain unchanged. 
Ref. \cite{ziel} attempted to put a more restrictive bound on
the $\rho\pi$ coupling of the $1^{-+}$ by using the assumption that the
state only decays to $\rho\pi$ and $\eta\pi$. As we shall see in the
next paragraph,
this is not a good assumption. 
Also, a limit of $\Gamma(\rhohat^+\rightarrow
\gamma\pi^+)\; BR(\rhohat^+\rightarrow\pi^+\eta)\lapprox 25$ keV has been
derived \cite{ziel}, which we translate within the assumption of VDM coupling of
the photon to
$\Gamma(\rhohat\rightarrow\rho\pi)\; BR(\rhohat\rightarrow\pi\eta)\lapprox
6$ MeV, using the equations in ref. \cite{ziel}. This is not very
restrictive, and we urge current experiments at BNL and VES which
sample $\eta\pi$ and $\rho\pi$ couplings at various invariant masses to derive a more restrictive
bound, since suppression of $\eta\pi$ is expected from symmetrization
selection rules \cite{tech}. 


 
Although evidence for a $1^{-+}$ isovector signal at 1.4 GeV  has
been reported in $\eta\pi$ by VES and BNL \cite{ves93,bnletapi}, 
no indication for of $\rhohat$ has been found in $\eta\pi$. This, together with
weak evidence for $\eta^{'}\pi$ coupling at
VES \cite{ves93} suggest that the ordering\footnote{
Assuming that the number of events is proportional to
$\Gamma(X\rightarrow\rho\pi)\; BR(X\rightarrow\eta^{(')}\pi)$, we make the rough estimates 
$BR(\rhohat\rightarrow\eta\pi)\lapprox 1\%$ from BNL and VES data
\protect\cite{ves93,bnletapi}
and $BR(\rhohat\rightarrow\eta^{'}\pi)\sim 3\%$ from VES data
\protect\cite{ves93}. If the structure in $\eta^{'}\pi$ is not taken to
be due to the $\rhohat$, we have the upper bound $BR(\rhohat\rightarrow\eta^{'}\pi)\lapprox 3\%$.} of partial widths for
$\rhohat$ is $\eta\pi < \eta^{'}\pi < \rho\pi$. 
If the $\rho\pi$
branching ratio is $20\pm 2\;\%$, as calculated above, the
question arises what the dominant modes of $\rhohat$ are. Since BNL
indicated that $b_1$ and $f_1$ exchange may be important \cite{bnl97},
as previously mentioned, decays to $b_1\pi$ and $f_1\pi$ may be dominant.

Is $\rhohat$ dominantly a hybrid, four--quark state or glueball?

The pure $SU(3)$ lattice gauge theory $1^{-+}$ glueball is predicted at mass
$\leq 4.1$ GeV with high confidence \cite{michael96}. However,
``there is no evidence for any [exotics] of mass less than 3 GeV'' \cite{michael96}.
This establishes that the $1^{-+}$ glueball is distant in mass relative to
the $\sim 2$ GeV mass region we are interested in. A preliminary
quenched lattice study found small $0^{++}$ glueball mixing with
pseudoscalar states \cite{weingarten}, indicative of weak glueball mixing. We henceforth
consider the $1^{-+}$ to be a linear combination of a
hybrid and four--quark state. 

The four--quark interpretation is disfavoured by the fact that no
stable configurations with $J^{PC}=1^{-+}$ have been predicted in 
models \cite{semay}, and that the total width is expected to be substantial
due to a lack of inhibition for four quarks to ``fall apart'' to meson final states.
If $\rhohat$ is a four--quark state, we expect at least two isovector
four--quark states in the same mass region \cite{semay}, since both
$(QQ)_{I=1}(\bar{Q}\bar{Q})_{I=0}$ and
$(QQ)_{I=0}(\bar{Q}\bar{Q})_{I=1}$
are independent degrees of freedom. 50\%--50\% linear combinations yield
two charge conjugation eigenstates with opposite  C--parity \cite{semay}. Thus if
$\rhohat$ is a four--quark state, we expect evidence for an isovector
$1^{--}$ at $\sim 1.6$ GeV. 

The molecular interpretation is questionable because $\rhohat$ is 
far from the $\rho\pi$ threshold it has been detected in. To eliminate
this interpretation, lack of coupling of $\rhohat$ to threshold
channels\footnote{$\rho\omega,\; \eta(1440)\pi,\; f_1(1420)\pi,\;
f_1(1510)\pi, f_2^{'}(1525)\pi$ and $f_2(1430)\pi$ are near the 1.6
GeV threshold.}
like $\rho(1450)\pi$ should be established.

On the other hand, all current information is consistent with the
hybrid interpretation:

\begin{enumerate}

\item The connected decays of exotic hybrid mesons to $\eta\pi$ and
$\eta^{'}\pi$ are suppressed by 
symmetrization selection rules \cite{page97sel1}, which is consistent
with the non--observation of the $\rhohat$ in $\eta\pi$ and the weak
$\eta^{'}\pi$ \cite{ves93,bnletapi}. Isovector four--quark state decay
to $\eta\pi$ and $\eta^{'}\pi$ 
are not necessarily suppressed by symmetrization selection rules,
although in certain configurations they can be \cite{page97sel1}. 
Although this does not rule out the four--quark interpretation, it
is consistent with $\hat{\rho}(1600)$ being a hybrid meson.

\item Single OZI forbidden diagrams can lead to flavour SU(3)
singlet production, since the pair created out of the vacuum is a
flavour singlet within SU(3) symmetry. This enhances $\eta^{'}\pi$ relative to $\eta\pi$ if
we neglect phase space, since
the $\eta^{'}$ has a large SU(3) singlet component \cite{page97sel1,lipkin1}. 
This is also confirmed by QCD sum rules calculations (as discussed in
ref. \cite{page97sel1}). All of this is consistent with the claim from VES
that $\eta^{'}\pi > \eta\pi$ when phase space is removed. 
This is not the expectation for a four--quark state \cite{page97sel1,lipkin1}. 

\item The small $\rho\pi$ coupling $BR(\rhohat\rightarrow
\rho\pi) \approx 20\pm 2\;\%$
previously extracted from the data is
consistent with model predictions for a hybrid meson\footnote{Amongst flux--tube models,
the IKP model gives $BR(\rhohat\rightarrow \rho\pi) = 10\%$ while the
PSS model gives 20\% \protect\cite{models}.} \cite{models}, where the $\rho\pi$
coupling is suppressed due to a selection rule \cite{page97sel2}.

\end{enumerate}

\noindent We list promising unexplored production mechanisms for $\rhohat$:

\begin{enumerate}


\item $p\bar{p}$ in the dominant $^1S_0$ configuration can annihilate
to $\rhohat\pi$ in P--wave. The decay $\rhohat\rightarrow\rho\pi\rightarrow (\pi^+\pi^-)\pi$ can be
accessed at Crystal Barrel.

\item Photoproduction of $1^{-+}$ has provisionally been approved at TJNAF
\cite{cebafb1pi} and since diffractive exchange has $C=+$ and is
electrically neutral, the production of $1^{-+}$ can only be via meson
exchange, particularly $\pi,\; \omega$ and $\rho$ exchange.

\end{enumerate}

\noindent $p\bar{p}$ central production at WA102, FNAL, RHIC and COMPASS and gluon jets at LEP2 can
also produce $\rhohat$.

We sketch further production mechanisms for exotics. 

{\it $\psi$ radiative decay:} The expectation is that $\Gamma$(hybrid$\rightarrow gg)\sim \alpha_S$ and
that the production of hybrid mesons in $\psi$ radiative decay should
be substantial\footnote{This is has been confirmed in a model calculation for
$J^{PC}=0^{-+}$ hybrids \protect\cite{page97jpsi}.}. BES should search for the
isoscalar partner of $\rhohat$ particularly in $\omega\omega,\;\rho\rho,\;
K^{\ast}K$ and $a_2\pi$.  

{\it Diffractive photoproduction:} Na\"{\i}ve estimates suggest that hybrid meson diffractive production
by incoming mesons should be enhanced above that of
glueballs, conventional mesons and four--quark states, due to the
presence $Q\bar{Q}$ {\it and} glue at the production vertex. This is
supported by the recent observation by VES of the hybrid meson candidate
\cite{models} $\pi(1800)$ in diffractive production. VES 
notes that ``the wave $J^{P}=0^-$ dominates at low'' $t<0.08$ GeV$^{2}$, indicating
diffraction \cite{ves}. We hence expect the diffractive production of
neutral exotic $2^{+-}$ hybrids in photoproduction, in analogy to the
diffractive process $\pi N\rightarrow a_1 N$ for which there is
experimental indications. Due to s--channel helicity conservation,
neutral exotic $0^{+-}$ can only be produced in electroproduction.

In conclusion, significant experimental and theoretical progress
on a $J^{PC}$ exotic state has been made. Further progress is expected
from searches in other decay channels and for other exotics in various production mechanisms.

\vspace{.5cm}

Helpful discussions with S.-U. Chung, J. Manak, 
D. Toussaint, D. Weygand and A. Zaitsev are acknowledged.
I acknowledge a Lindemann Fellowship from the English Speaking Union.

\appendix

\end{document}